\documentclass[american,onecolumn]{revtex4}
\usepackage{graphicx}
\usepackage{amssymb}

\usepackage{babel}

\begin{document}

\title{On the origin of inflation }

\author{Torsten Asselmeyer-Maluga}

\email{torsten.asselmeyer-maluga@dlr.de}

\affiliation{German Aerospace Center (DLR), Rutherfordstr. 2, 12489 Berlin, Germany}

\author{Jerzy Kr\'ol}

\email{iriking@wp.pl}

\affiliation{University of Silesia, Institute of Physics, ul. Uniwesytecka 4,
40-007 Katowice, Poland}
\begin{abstract}
In this paper we discuss a space-time having the topology of $S^{3}\times\mathbb{R}$
but with different smoothness structure. This space-time is not a
global hyperbolic space-time. Especially we obtain a time line with
a topology change of the space from the 3-sphere to a homology 3-sphere
and back but without a topology-change of the space-time. Among the
infinite possible smoothness structures of this space-time, we choose
a homology 3-sphere with hyperbolic geometry admitting a homogenous
metric. Then the topology change can be described by a time-dependent
curvature parameter $k$ changing from $k=+1$ to $k=-1$ and back.
The solution of the Friedman equation for dust matter $(p=0)$ after
inserting this function shows an exponential growing which is typical
for inflation. In contrast to other inflation models, this process
stops after a finite time.
\end{abstract}
\pacs{04.20.Gz, 98.80.Cq, 02.40.Ma}
\keywords{smoothness structures of space-time, spatial topology change, inflation}
\maketitle
Because of the influx of observational data, recent years have witnessed
enormous advances in our understanding of the early universe. To interpret
the present data, it is sufficient to work in a regime in which space-time
can be taken to be a smooth continuum as in general relativity, setting
aside fundamental questions involving the deep Planck regime. However,
for a complete conceptual understanding as well as interpretation
of the future, more refined data, these long-standing issues will
have to be faced squarely. As an example, can one show from first
principles that the smooth space-time of general relativity is valid
at the onset of inflation? In this paper we will focus mainly on this
question about the origin of inflation. Inflation is today the main
theoretical framework that describes the early Universe and that can
account for the present observational data \cite{WMAP-7-years}. In
thirty years of existence \cite{Guth1981,Linde1982}, inflation has
survived, in contrast with earlier competitors, the tremendous improvement
of cosmological data. In particular, the fluctuations of the Cosmic
Microwave Background (CMB) had not yet been measured when inflation
was invented, whereas they give us today a remarkable picture of the
cosmological perturbations in the early Universe. In nearly all known
models, the inflation period is caused by one or more scalar field(s)\cite{InflationBook}.
A different approach is Loop quantum cosmology containing also the
inflation scenario in special cases \cite{AshBojLew:03,AshPawSin:06}.

In this paper we try to derive an inflationary phase from first principles.
The analysis of the WMAP data seems to imply that our universe is
a compact 3-manifold with a slightly positive curvature \cite{WMAPcompactSpace2008}.
Therefore we choose the topology of the space-time to be $S^{3}\times\mathbb{R}$.
Clearly this space-time admits a Lorentz metric (given by the topological
condition to admit a non-vanishing vector field normal to $S^{3}$).
But we weaken the condition of global hyperbolicity otherwise it induces
a diffeomorphism\cite{BernalSanchez2003,BernalSanchez2006} to $S^{3}\times\mathbb{R}$.
The space-time has the topology of $S^{3}\times\mathbb{R}$ and is
smoothable (i.e. a smooth 4-manifold) \cite{Qui:82} but (we assume)
is not diffeomorphic to $S^{3}\times\mathbb{R}$. What does it mean?
Every manifold is defined by a collection of charts, the atlas, including
also the transition functions between the charts. From the physical
point of view, charts are the reference frames. The transition functions
define the structure of the manifold, i.e. transition functions are
homeomorphisms (topological manifold) or diffeomorphisms (smooth manifold).
Two (smooth) atlases are compatible (or equivalent) if their union
is a (smooth) atlas again. The equivalence class (the maximal atlas)
is called a differential structure%
\footnote{A smooth atlas defines a smoothness structure.%
}. In dimension smaller than $4$, there is only one differential structure,
i.e. the topology of these manifolds define uniquely its smoothness
properties. In contrast, beginning with dimension $4$ there is the
possibility of more than one differential structure. But 4-manifolds
are really special: here there are many examples of 4-manifolds with
infinite many differential structures (countable for compact and uncountable
for non-compact 4-manifolds including $\mathbb{R}^{4}$). Among these
differential structures there is one exceptional, the standard differential
structure. We will illustrate these standard structure for our space-time
$S^{3}\times\mathbb{R}$. The 3-sphere $S^{3}$ has an unique differential
structure (the standard differential structure) which extends to $S^{3}\times\mathbb{R}$.
All other differential structures (also called misleadingly ''exotic
smoothness structures'') can never split smoothly into $S^{3}\times\mathbb{R}$.
We denote it by $S^{3}\times_{\theta}\mathbb{R}$. Our main hypothesis
is now:\\
\textbf{Main hypothesis}: \emph{The space-time has the topology
$S^{3}\times\mathbb{R}$ but the differential structure $S^{3}\times_{\theta}\mathbb{R}$.}\\
In \cite{Fre:79} the first $S^{3}\times_{\theta}\mathbb{R}$
was constructed and we will use this construction here. One starts
with a homology 3-sphere $P$, i.e. a compact 3-manifold $P$ with
the same homology as the 3-sphere but non-trivial fundamental group,
see \cite{Bre:93}. The Poincare sphere is one example of a homology
3-manifold. Now we consider the 4-manifold $P\times[0,1]$ with the
same fundamental group $\pi_{1}(P\times[0,1])=\pi_{1}(P)$. By a special
procedure (the plus construction, see \cite{Mil:61,Ros:94}), one
can ''kill'' the fundamental group $\pi_{1}(P)$ in the interior of
$P\times[0,1]$. This procedure will result in a 4-manifold $W$ with
boundary $\partial W=-P\sqcup S^{3}$ ($-P$ is $P$ with opposite
orientation), a so-called cobordism between $P$ and $S^{3}$. The
gluing $-W\cup_{P}W$ along $P$ with the boundary $\partial(-W\cup_{P}W)=-S^{3}\sqcup S^{3}$
defines one piece of the exotic $S^{3}\times_{\Theta}\mathbb{R}$.
The whole construction can be extended to both directions to get the
desired exotic $S^{3}\times_{\Theta}\mathbb{R}$ (see \cite{Fre:79,Kir:89}
for the details). There is one critical point in the construction:
the 4-manifold $W$ is not a smooth manifold. As Freedman \cite{Fre:82}
showed, the 4-manifold $W$ always exists topologically but, by a
result of Gompf \cite{Gom:89} (using Donaldson \cite{Don:83}), not
smoothly (i.e. it does not exists as a smooth 4-manifold). The 4-manifold
$-W\cup_{P}W$ is also non-smoothable and we will get a smoothness
structure only for the whole non-compact $S^{3}\times_{\Theta}\mathbb{R}$
(see \cite{Qui:82}). But $S^{3}\times_{\Theta}\mathbb{R}$ contains
$-W\cup_{P}W$ with the smooth cross section $P$. From the physical
point of view we interpret $-W\cup_{P}W$ as a time line of a cosmos
starting as 3-sphere changing to the homology 3-sphere $P$ and changing
back to the 3-sphere. But this process is part of every exotic smoothness
structure $S^{3}\times_{\Theta}\mathbb{R}$, i.e. we obtain the mathematical
fact\\
\textbf{Fact}\emph{: In the space-time $S^{3}\times_{\Theta}\mathbb{R}$
we have a change of the spatial topology from the 3-sphere to some
homology 3-sphere $P$ and back but without changing the topology
of the space-time.}\\
Now we have to discuss the choice of the homology 3-sphere $P$.
At first, usually every homology 3-sphere is the boundary of a topological,
contractable 4-manifold \cite{Fre:82} but this homology 3-sphere
$P$ never bounds a \textbf{smooth}, contractable 4-manifold. Secondly,
every homology 3-sphere can be constructed by using a knot \cite{Rol:76}.
One starts with the complement $S^{3}\setminus\left(D^{2}\times K\right)$
of a knot $K$ (a smooth embedding $S^{1}\to S^{3}$) and glue in
a solid torus $D^{2}\times S^{1}$ using a special map (a $\pm1$
Dehn twist). The resulting 3-manifold $\Sigma(K)$ is a homoloy 3-sphere.
For instance the trefoil knot $3_{1}$ (in Rolfsen notation \cite{Rol:76})
generates the Poincare sphere by this method (with $-1$ Dehn twist). 

Our model\emph{ }$S^{3}\times_{\Theta}\mathbb{R}$ starts with a 3-sphere
as spatial topology. Now we will be using a powerful tool for the
following argumentation, Thurstons geometrization conjecture \cite{Thu:97}
proved by Perelman \cite{Per:02,Per:03.1,Per:03.2}. According to
this theory, only the 3-sphere and the Poincare sphere carry a homogenous
metric of constant positive scalar curvature (spherical geometry or
Bianchi IX model) among all homology 3-spheres. Also other homology
3-spheres%
\footnote{The homology 3-sphere must be irreduzible and therefore generated
by an irreduzible knot, i.e. a knot not splittable in a sum of two
non-trivial knots.%
} are able to admit a homogenous metrics. There is a close relation
between Thurstons geometrization theory and Bianchi models in cosmology
\cite{Anderson,Andersson}. Most of the (irredussible) homology 3-spheres
have a hyperbolic geometry (Bianchi V model), i.e. a homogenous metric
of negative curvature. Here we will choose such a hyperbolic homology
3-sphere. As an example we choose the knot $8_{10}$ leading to the
hyperbolic homology 3-sphere $\Sigma(8_{10})$. So we will give an
overview of our assumptions:
\begin{enumerate}
\item The space-time is $S^{3}\times_{\Theta}\mathbb{R}$ (with\emph{ }topology
$S^{3}\times\mathbb{R}$) containing a homology 3-sphere $P$ (as
cross section).
\item This homology 3-sphere $P$ is a hyperbolic 3-manifold (with negative
scalar curvature).
\end{enumerate}
According to the mathematical fact above, the 3-sphere is changed
to $P$ and back in $S^{3}\times_{\Theta}\mathbb{R}$, i.e. we obtain
a topology change of the spatial cosmos. Now we will study the geometry
and topology changing process more carefully. Let us consider the
Robertson-Walker metric (with $c=1$)\[
ds^{2}=dt^{2}-a(t)^{2}h_{ik}dx^{i}dx^{k}\]
with the scaling function $a(t)$. At first we assume a space-time
$S^{3}\times\mathbb{R}$ with increasing function $a(t)$ fulfilling
the Friedman equations\begin{eqnarray}
\left(\frac{\dot{a}(t)}{a(t)}\right)^{2}+\frac{k}{a(t)^{2}} & = & \kappa\frac{\rho}{3}\label{eq:friedman-1}\\
2\left(\frac{\ddot{a}(t)}{a(t)}\right)+\left(\frac{\dot{a}(t)}{a(t)}\right)^{2}+\frac{k}{a(t)^{2}} & = & -\kappa p\label{eq:friedman-2}\end{eqnarray}
derived from Einsteins equation\begin{equation}
R_{\mu\nu}-\frac{1}{2}g_{\mu\nu}R=\kappa T_{\mu\nu}\label{eq:Einstein-equation}\end{equation}
with the gravitational constant $\kappa$ and the energy-momentum
tensor of a perfect fluid\begin{equation}
T_{\mu\nu}=(\rho+p)u_{\mu}u_{\nu}-pg_{\mu\nu}\label{eq:perfect-fluid-EM-tensor}\end{equation}
with the (time-dependent) energy density $\rho$ and the (time-dependent)
pressure $p$. The spatial cosmos has the scalar curvature $^{3}R$\[
^{3}R=\frac{k}{a^{2}}\]
from the 3-metric $h_{ik}$ and we obtain the 4-dimensional scalar
curvature $R$\begin{equation}
R=\frac{6}{a^{2}}\left(\ddot{a}\cdot a+\dot{a}^{2}+k\right)\:.\label{eq:4-dim-scalar-curvature}\end{equation}
Let us consider the model $S^{3}\times\mathbb{R}$ with positive spatial
curvature $k=+1$. In case of dust matter ($p=0$) only, one obtains
a closed universe. Now we consider our model of an exotic $S^{3}\times_{\Theta}\mathbb{R}$.
As explained above, the foliation of $S^{3}\times_{\Theta}\mathbb{R}$
must contain a hyperbolic homology 3-sphere $P=\Sigma(8_{10})$ (with
negative scalar curvature). But then we have a transition from a space
with positive curvature to a space with negative curvature and back.
To model this behavior, we consider a time-dependent parameter $k(t)$
in the curvature \[
^{3}R(t)=\frac{k(t)}{a^{2}}\]
with the following conditions:
\begin{enumerate}
\item The change of the geometry from spherical $k>0$ to hyperbolic $k<0$
happens at $t_{0}$,
\item $k(t)>0$ for $t\ll t_{0}$ and $t\gg t_{0}$.
\end{enumerate}
Furthermore, $k(t)$ for $t\ll t_{0}$ must be larger than $k(t)$
for $t\gg t_{0}$. The change of the topology is an abrupt process
which can be modeled by a $\tanh$ function. Putting all these conditions
together we choose

\[
k(t)=\rho_{0}\exp\left(-\tanh(\zeta(t-t_{0}))\right)-\zeta^{2}\left(1-\tanh^{2}(\zeta(t-t_{0}))\right)^{2}\exp\left(2\cdot\tanh(\zeta(t-t_{0}))\right)\:.\]
\begin{figure}
\includegraphics[scale=0.4]{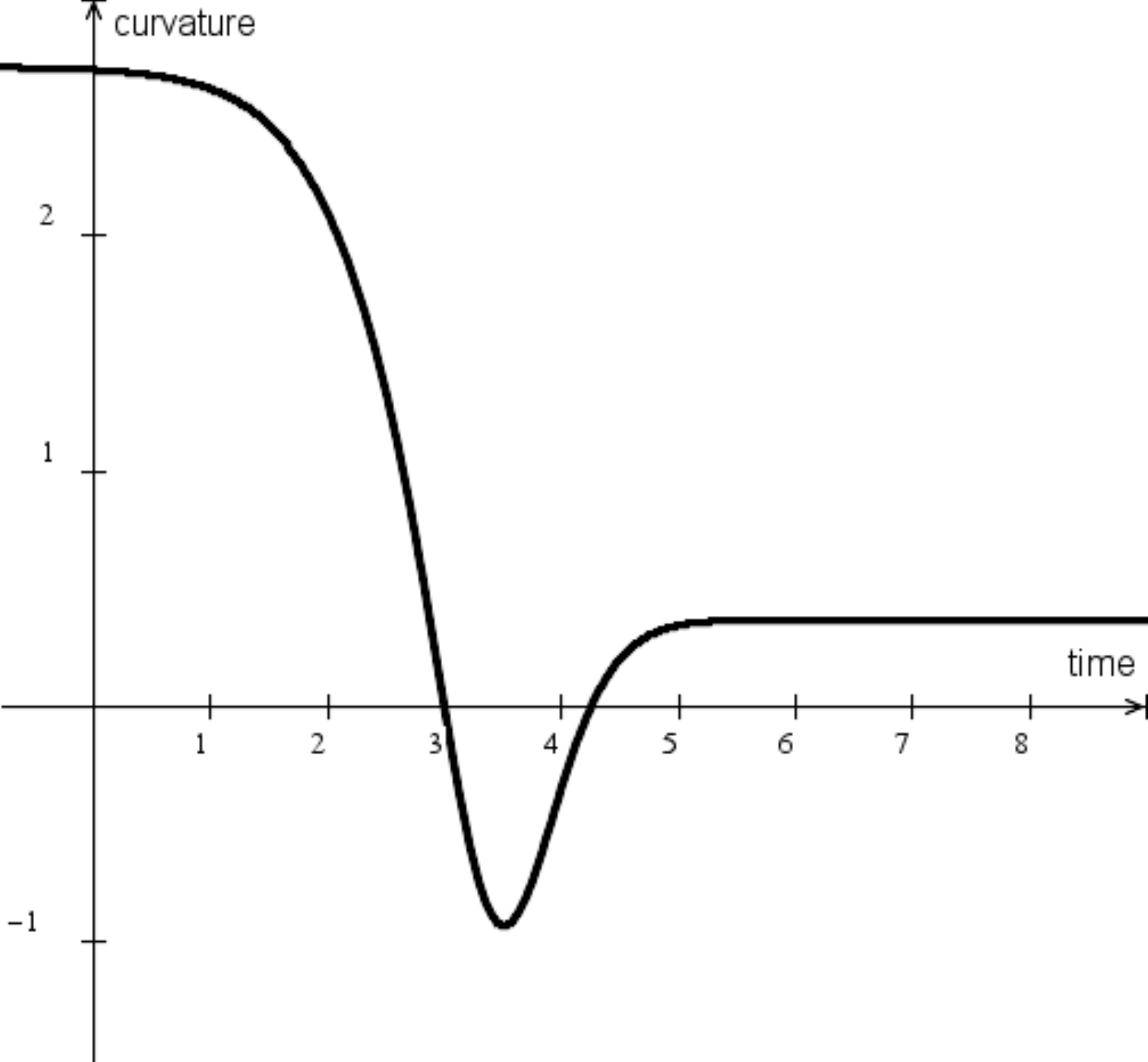} \qquad\includegraphics[scale=0.4]{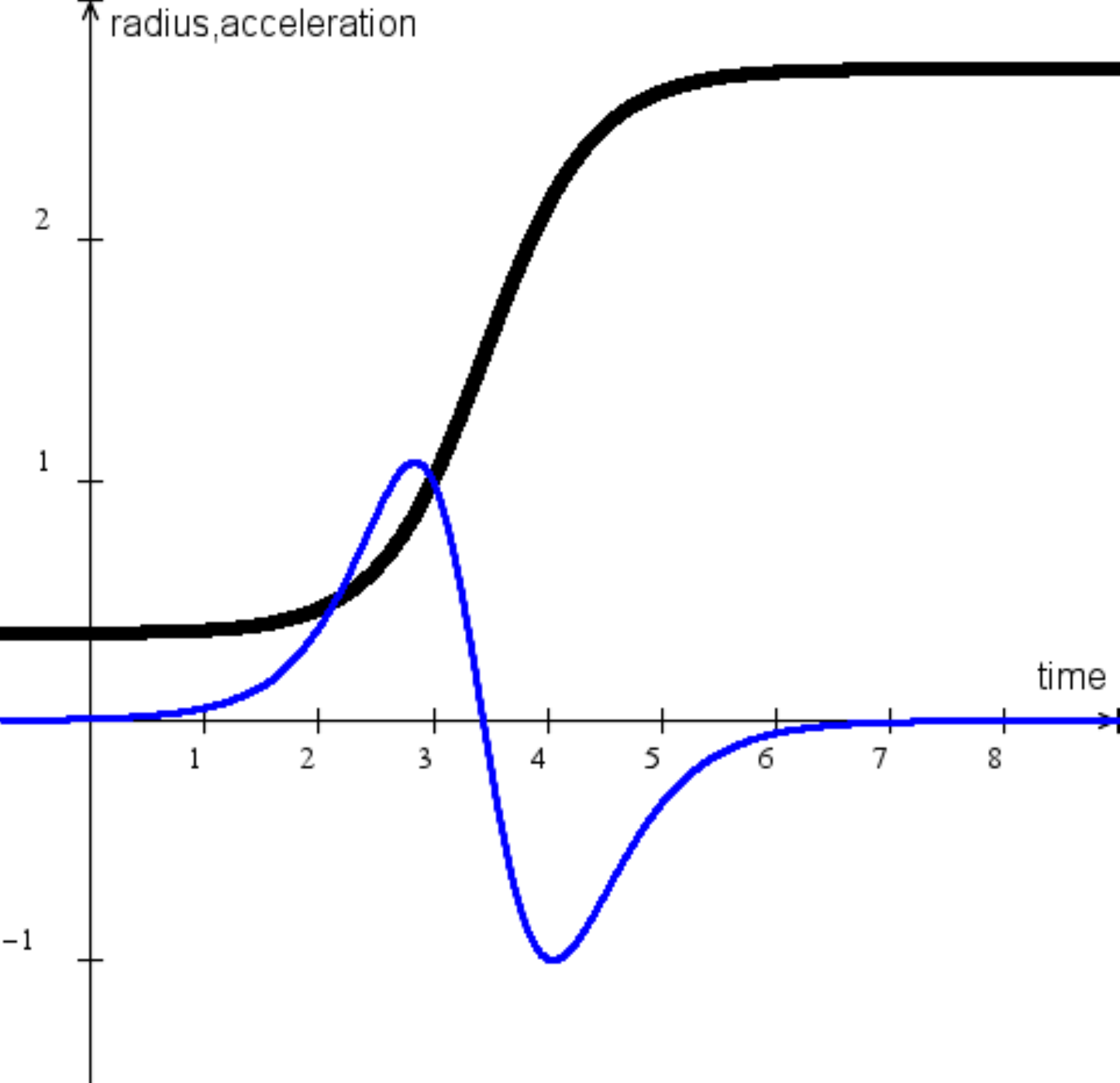}

\caption{time-dependent curvature $k(t)$ with $t_{0}=3,\zeta=1,\rho_{0}=1$(left
fig.) and the solution $a(t)$ (thick curve) as well the acceleration
$\ddot{a}(t)$ (right fig.)\label{fig:time-dependent-curvature-solution}}

\end{figure}
This function is plotted in the left figure of Fig. \ref{fig:time-dependent-curvature-solution}
for special parameters confirming the conditions above. Now we consider
the special Friedman equation\[
\left(\frac{\dot{a}(t)}{a(t)}\right)^{2}+\frac{k(t)}{a(t)^{2}}=\frac{\rho_{0}}{a(t)^{3}}\]
with dust matter $(p=0)$ (having the scaling behavior $\rho\sim a^{-3}$)
by inserting the time-dependent $k(t)$ above. The solution $a(t)$
of this equation is given by\[
a(t)=\exp\left(\tanh(\zeta(t-t_{0}))\right)\]
also visualized in the right figure of Fig. \ref{fig:time-dependent-curvature-solution}.
Especially we obtain \emph{a positive acceleration} $\ddot{a}(t)$
in some time interval. The \emph{growing rate is exponential} as required
by inflation. Furthermore the acceleration $\ddot{a}(t)$ is also
negative or the \emph{inflation process stops}. Finally the topology
change of the space\[
\mbox{spherical 3-sphere}\longrightarrow\mbox{hyperbolic homology 3-sphere}\longrightarrow\mbox{spherical 3-sphere}\]
produces an exponential growing rate of $a(t)$, also called inflation.
But in contrast to the usual inflation models, we derive this behavior
from first principles using the space-time $S^{3}\times_{\theta}\mathbb{R}$
(with a non-standard differential structure). Furthermore \emph{in
our inflation model, the exponential growing stops, i.e. our inflation
is not eternal}. 

But what is about the inflation without quantum effects? Fortunately,
there is growing evidence that the differential structures constructed
above (i.e. exotic smoothness in dimension 4) is directly related
to quantum gravitational effects \cite{Ass2010,Duston2010}. We will
further investigate this interesting direction in our future work.


\end{document}